\documentclass[apj]{emulateapj}
\usepackage{natbib}
\usepackage{amssymb}
\usepackage{amsmath}
\usepackage{graphicx}
\usepackage{color}

\def \Ni {$^{56}$Ni}

\def \Lb {L_{\rm bol} }

\def \Eexp {E_{\rm exp}}
\def \Eenv {E_{\rm env}}

\def \Mej{M_{\rm ej}}
\def \vej{v_{\rm ej}}
\def \venv{v_{\rm env}}
\def \Menv{M_{\rm env}}
\def \Mi{M_{\rm ZAMS}}
\def \Renv{\overline{R}_{\rm env}}

\def \Q {Q_{\rm Ni} }

\def \Et {ET}
\def \sref {\S \ref}

\begin{document}
\newcommand{\vv}{\textrm{v}}
\title{Type II supernovae progenitor and ejecta properties from the total emitted light, $\Et$}
\author{Tomer Shussman\altaffilmark{1}, Ehud Nakar\altaffilmark{1}, Roni Waldman \altaffilmark{2,3},  Boaz Katz\altaffilmark{3}}
\altaffiltext{1}{The Raymond and Beverly Sackler School of Physics and
Astronomy, Tel Aviv University, Tel Aviv 69978, Israel}
\altaffiltext{2}{Racah Institute of Physics, The Hebrew University, Jerusalem 91904, Israel}
\altaffiltext{3}{Particle Physics \& Astrophysics Dept., Weizmann Institute of Science, Rehovot 76100, Israel}

\begin{abstract}
It was recently shown that the bolometric light curves of type II supernovae (SNe) allow an accurate and robust measurement of the product of the radiation energy in the ejecta, $E_r$, and the time since the explosion, $t$, at early phases ($t\lesssim 10d$) of the homologous expansion. This observable, denoted here $\Et \equiv E_rt$ is constant during that time and depends only on the progenitor structure and explosion energy. We use a 1D hydrodynamic code to find $\Et$ of simulated explosions of 145 red supergiant progenitors obtained using the stellar evolution code MESA, and relate this observable to the properties of the progenitor and the explosion energy. We show that $\Et$ probes only the properties of the envelope (velocity, mass and initial structure), similarly to other observables that rely on the photospheric phase emission. Nevertheless, for explosions where the envelope dominates the ejected mass, $M_{\rm env}/M_{\rm ej} \gtrsim 0.6$, $ET$ is directly related to the explosion energy $E_{\rm exp}$ and ejected mass $M_{\rm ej}$ through the relation $ET \approx 0.15 E_{\rm exp}^{1/2} R_* M_{\rm ej}^{1/2}$, where $R_*$ is the progenitor radius, to an accuracy better than $30\%$. We also provide relations between $\Et$ and the envelope properties that are accurate (to within 20\%) for all the progenitors in our sample, including those that lost most of their envelope. We show that when the envelope velocity can be reasonably measured by line shifts in observed spectra, the envelope is directly constrained from the bolometric light curve (independent of $\Eexp$). We use that to compare observations of 11 SNe with measured $\Et$ and envelope velocity to our sample of numerical progenitors. This comparison suggests that many SNe progenitors have radii that are $\lesssim 500~R_\odot$. In the framework of our simulations this indicates, most likely, a rather high value of the mixing length parameter.
\end{abstract}

\section{Introduction}\label{sec:Introduction}
The most common type of supernovae (SNe) are type II supernovae (supernovae with spectra showing significant amounts of hydrogen) and yet it is still unclear how these explosions work. While several lines of evidence indicate that the explosion is associated with the latest stages of stellar evolution of some massive stars and the collapse of their cores, the process in which the envelope is energetically ejected is poorly understood. 

Recent extensive surveys and followup efforts are leading to the accumulation of large samples of well observed type II SNe. Given the diversity in the observed properties (e.g. peak luminosity), statistical inferences are likely to play an important role in improving our understanding of these events. Most observations of SNe in general, and type II in particular, consist of spectra and light curves of visible light acquired throughout weeks and months following the explosion. In order to connect the observed emission to the physical characteristics of the explosion and the progenitor star, studies usually either employ detailed radiation transfer calculations or use simplistic analytic models. While the former are more trustworthy, they are difficult to apply to the growing large samples of observed SNe, especially given the uncertainty in the late phase structure of the massive stars which are likely the progenitors. The application of simplistic analytic models is useful but may lead to crude errors in estimates of the interesting properties such as explosion energy, mass and radius of the progenitor. 

Recently, we showed \citep{Nakar15,Katz13} that the bolometric light curve can be used to extract information on the explosion which circumvents the difficulty of radiation transfer by direct use of energy conservation. For that we defined a time-weighted integrated luminosity (with the \Ni\ contribution subtracted)  which is directly observable. We denoted it $\Et$ and in \cite{Nakar15} measured it for a sample of 13 SNe. $\Et$ is set by the energetics of the explosion and the structure of the progenitor and it provides a direct relation between observations and explosion models which is simple, analytic and precise at the same time. 

The purpose of this paper is to use hydrodynamic simulations to study the relation between this measurable quantity and useful parameters of the progenitor and explosion. To do that we calculate ET for a large set of 145 red supergiant models that are obtained by the stellar evolution code MESA \citep{2011ApJS..192....3P,2013ApJS..208....4P,2015ApJS..220...15P} by varying progenitor parameters on the main sequence (initial mass, metallicity and rotation) and an evolution parameter (mixing length parameter).  

The paper is organized as follows. In section \sref{sec:Et} we briefly repeat the arguments of \citet{Nakar15} and \citet{Katz13}, and explain how the total radiation energy can be measured and how exactly it is connected to the hydrodynamic properties of the explosion. In section \sref{sec:Cet} we describe the set of numerical progenitors we use and present the relations we find between $\Et$ and fundamental properties of the progenitor and explosion energy. In section \sref{sec:spectral} we use observations of SNe with measured ET and envelope velocity to constrain the envelope mass and radius, and SNe with measured ET, envelope velocity and pre-explosion radius to constrain the explosion energy.

\section{$\Et$ is measurable, independent of radiation transfer, and scales as $vMR$}\label{sec:Et}
We first briefly repeat the arguments of \citet{Nakar15} and \citet{Katz13}. During the time span of about 1 to 10 days after the explosion, the expansion is homologous and the radiation is almost entirely trapped within the ejecta. In addition, the contribution of energy from the decay of \Ni\ is negligible. The total energy in radiation $E_r$, originating from the explosion shock wave, decreases with time adiabatically due to the work it does on the expanding ejecta. To a very good approximation it decreases in this phase as $E_r\propto1/t$ where $t$ is the time since explosion and thus we define 
\begin{equation}\label{eq:Etdef1}
\Et \equiv E_rt = {\rm const}~~~~~~\rm  1\rm{d} \lesssim t\lesssim 10\rm{d}
\end{equation}
which is constant with time.
Since there is negligible diffusion during this time, the quantity $\Et$ is completely set by the hydrodynamic properties of the explosion and is  independent of the opacity of the ejecta. At later times diffusion becomes important and the trapped internal energy leaks out of the ejecta  gradually to generate the observed luminosity. Thus, as shown by \cite{Nakar15}, by integrating over the observed bolometric luminosity (multiplied by $t$ to compensate for adiabatic losses), and removing the contribution from $^{56}$Ni, one can directly extract $\Et$ from observations.    

To show that formally we start from the equation that describes the radiation energy which is trapped in the ejecta during the homologous phase:
\begin{equation}\label{eq:dE_dt}
	\frac{dE_r(t)}{dt}=-\frac{E_r(t)}{t}+\Q(t)-\Lb (t) .
\end{equation} 
Here, $\Q(t)$ is the energy injection rate from \Ni\ decay and $\Lb(t)$ is the bolometric luminosity. The term $-E_r/t$ is the total rate of adiabatic loss. After rearranging the equation, multiplying both sides by $t$ and integrating over $t$ in some interval $t_1<t'<t_2$ we obtain  

\begin{equation}\label{eq:Integral}
	\int^{t_2}_{t_1} t'\,\Lb(t')\,dt'=E_r(t_2)t_2 - E_r(t_1)t_1+\int^{t_2}_{t_1}t'\,\Q(t')\,dt.
\end{equation}

By choosing $t_1$ to be in the range $1\rm{d}\lesssim t\lesssim 10\rm{d}$, we can replace $E_r(t_1)t_1$ in equation \ref{eq:Integral} by $\Et$. By choosing $t_2$ that is large enough (typically $\gtrsim 120$d) so the diffusion time is shorter than the expansion time and the remaining radiation in the ejecta is negligible, we can neglect the term $E_r(t_2)t_2$. Finally, since the (time weighted) integrated luminosity at early times is negligible, we can extend the integration from $t=0$ without affecting the result. We therefore find

\begin{equation}\label{eq:Integral2}
	\Et=\int_{0}^{t_2\gtrsim 120\rm{d}} t'\,(\Lb(t') - \Q(t'))\,dt'.
\end{equation}
The second term on the RHS of equation \ref{eq:Integral2} can be calculated using 
\begin{equation}\label{eq:QNi}
\Q(t)=\frac{M_{Ni}}{M_{\odot}}(6.45 e^{-t_d/8.8}+1.45e^{-t_d/111.3})\times 10^{43}{\rm erg\,s^{-1}}
\end{equation}
where $t_d=t/$d and $M_{Ni}$ is the \Ni\ mass ejected in the explosion. $M_{Ni}$ can be accurately inferred from the amplitude of the bolometric luminosity at late times $120\rm{d}\lesssim t \lesssim 300\rm{d}$ where to a very good approximation 
\begin{equation}\label{eq:Nitail}
\Lb(t)=\Q(t)~~~~~ 120\rm{d} \lesssim t\lesssim 300\rm{d}
\end{equation}
(at later times $\gamma$-ray escape may become significant). The value of $\Et$ can thus be directly extracted from observations of type II SNe using equations \ref{eq:Integral2} - \ref{eq:Nitail}, as long as the bolometric luminosity is measured up to the \Ni\ tail phase (where $\Lb=\Q\propto e^{-t_d/111.3}$).

\begin{figure}[h!]
\includegraphics[width=\linewidth]{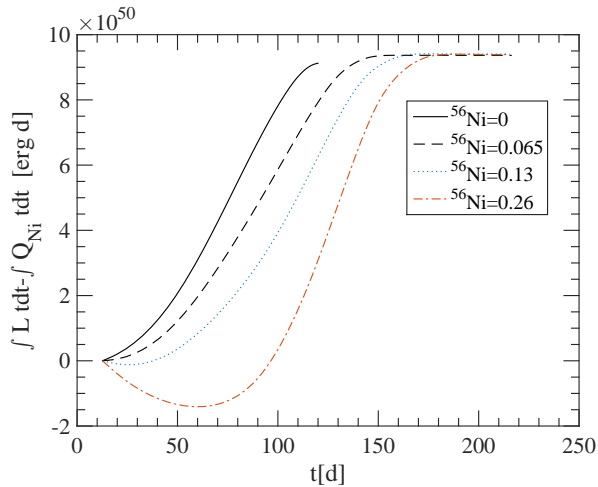}
\caption{Synthetic $\Et$ extracted from the radiation transfer simulations of a type II SNe in figure 2 of \citet{Kasen09}. The value of $\Et$ is the asymptotic value of the curve $\int \Lb t dt-\int \Q tdt$ shown as expressed in equations \ref{eq:Integral2} \& \ref{eq:QNi}. The values of \Ni\ were taken from \citet{Kasen09}. As can be seen, the value of $\Et$ is independent of the amount of \Ni\ to an accuracy of a few percents.
\label{fig:Kasen09}}
\end{figure}

We note that $\Et$ is also equal to the (time weighted) integrated 'cooling envelope luminosity' $L_e$ defined as the luminosity that would have been generated if there was no \Ni\ present in the ejecta
\begin{equation}\label{eq:ETLe}
\Et=\int_0^{\infty}L_e(t)t dt.
\end{equation}
Equation \ref{eq:ETLe} is useful for studying the different approximations in radiation transfer simulations were the \Ni\ can be artificially extracted. This is demonstrated in figure \ref{fig:Kasen09} using the results of radiation transfer simulations reported in \citep{Kasen09}.

We next derive the scaling we expect for $\Et$ with the progenitor radius, $R_*$, ejecta mass, $\Mej$, and total explosion energy, $\Eexp$. In this paper we ignore the inner collapsing parts of the progenitor and the initial thermal and gravitational energy which is negligible when considering the material at large radii ($r\gtrsim 10^{10}\rm cm$). The explosion is thus described as a shock wave traversing a cold standing star and a following expansion. While the shock is within the star, there is negligible diffusion and the thermal energy is dominated by radiation. Thus, for a set of progenitors with the same density profile\footnote{Two progenitors, denoted as '1' and '2', have the same density profile if $\rho_1(r_1)\frac{R_{*,1}^3}{M_1}=\rho_2(r_2)\frac{R_{*,2}^3}{M_2}$, where $\rho$ is density, $M$ is the progenitor mass and $r_1=r_2\frac{R_{*,1}}{R_{*,2}}$}, by the time the shock ends traversing the star and breaks out the radiation energy contained in the ejecta is $\propto \Eexp$ and the expansion time over which significant adiabatic losses take place is $\propto R_*/\vej \propto R_*\Eexp^{-1/2}\Mej^{1/2}$. Thus for progenitors with similar profiles $\Et$ scales as
\begin{equation}\label{eq:ETscaling}
\Et \propto \Eexp^{1/2}R_*\Mej^{1/2} \propto \frac{\Eexp R_*}{\vej} \propto \vej R_* \Mej,
\end{equation}
where 
\begin{equation}
\vej=\sqrt{2\Eexp/\Mej}=\sqrt{\frac{\int v(m)^2 dm}{\Mej}}
\end{equation}
is the (mass weighted) RMS velocity of the ejecta ($v(m)$ is the velocity of the mass element $dm$). However, since stars with different initial conditions (e.g., ZAMS mass, metallicity, rotation, binarity, etc.) have different density profiles before they explode, the coefficient in equation \ref{eq:ETscaling} for each progenitor structure is expected to be different. Below we calculate $\Et$ for a large set of progenitors and find the typical value of the  coefficient and how it varies between different progenitors. We also study which of the progenitor and explosion properties can be constrained best by measuring $\Et$.

\section{Numerical study of the scaling of $\Et$ with progenitor parameters}\label{sec:Cet}

\begin{figure}[t!]
\includegraphics[width=\linewidth]{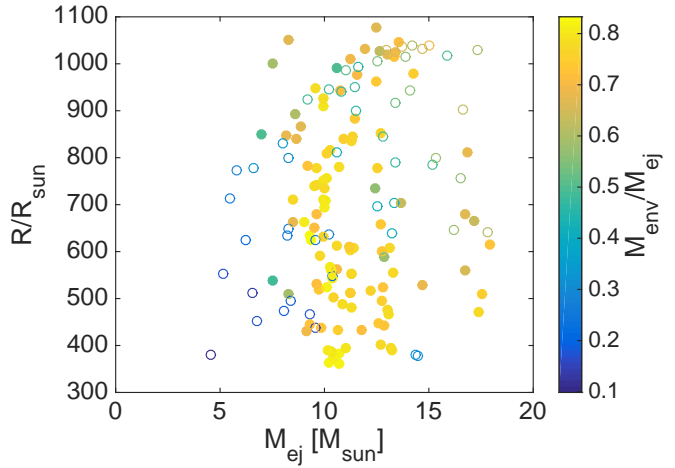}
\caption{Properties of the numerical progenitors calculated via MESA. Plotted are the radius versus the ejecta mass (the remnant mass is irrelevant for our analysis). Different colors represent different values of the ratio between the envelope and ejecta mass. The filled circles represent progenitors with $\Mi \leq 20M_{\odot}$ , which seems to compose most (or all) of SNe II-P \citep{Smartt15}. The empty circles represent progenitors with $\Mi >20M_{\odot}$
\label{fig:progenitors}}
\end{figure}

We have used the open source stellar evolution code MESA \citep{2011ApJS..192....3P,2013ApJS..208....4P,2015ApJS..220...15P} to calculate a large set of progenitors. Only single star evolution is considered (no binarity) and three initial conditions are varied - initial (ZAMS) mass (12-50 $M_\odot$), metallicity ($10^{-3}-1~{\rm z_\odot}$) and initial rotation (0-0.8 of breakup rotation rate). We have also varied one poorly constrained evolution parameter - the mixing length parameter (1.5-5). Altogether we calculate the stellar structure at the time of core collapse for 219 stars. Out of these 145 retain a hydrogen envelope at the time of explosion and are therefore considered here as plausible progenitors of type II SNe. Figure \ref{fig:progenitors} depicts some of the main properties of these 145 progenitors. The evolution parameters, together with the main properties at explosion are given in table 2. More details on the set of progenitors we use here are given in \cite{Shussman16}

In order to calculate $\Et$ for all the progenitors that retain hydrogen envelope, we excise the Si core from each progenitor (considered to be the explosion remnant), and explode it using a a simple 1D hydro-radiation code (see \citealt{Shussman16} for more details on the code). The explosion is induced by instantaneously releasing the explosion energy at the center of the ejecta. We then calculate $ET$, by extracting the asymptotic value of $E_rt$ at late times ($t=100$ d) using hydrodynamics alone, without allowing for radiation to diffuse. We have verified that the results remain unchanged when radiation transfer is included, in which case $\Et$ is calculated using equation \ref{eq:ETLe}. As explained above (equation \ref{eq:ETscaling}) for a given progenitor we expect $\Et\propto \Eexp ^{1/2}$. We find that this is indeed the case in our simulations (to within 1\%). Thus, as the dependence of $\Et$ on $\Eexp$ is known, we focus here on its dependence on the progenitor properties.

\begin{figure}[t!]
\includegraphics[width=\linewidth]{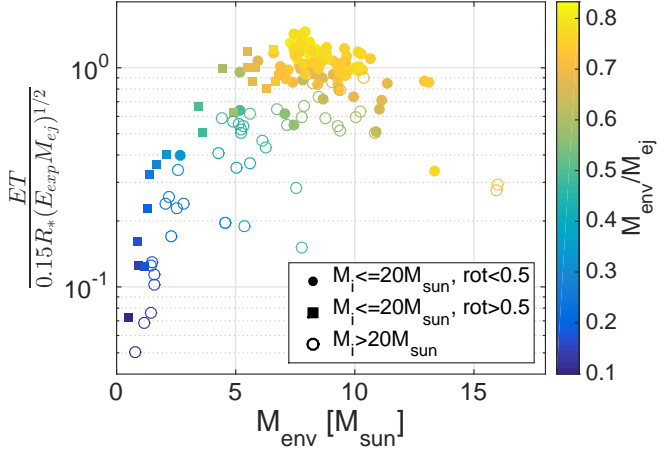}
\caption{Ratio of $\Et$ and the estimator $\Eexp^{1/2}R_*\Mej^{1/2}$ based on equation \ref{eq:ETscaling}. 
{\it Empty circles} mark progenitors with $\Mi>20M_\odot$. {\it Filled squares} mark fast rotating ($>0.5$ of the breakout rate) progenitors with $\Mi<20M_\odot$ and {\it filled circles} mark progenitors with $\Mi<20M_\odot$ that are not fast rotating. As can be seen, the simple scaling of equation \ref{eq:ETscaling} captures the changes in $\Et$ in progenitors with $\Menv/\Mej>0.6$ but fails in progenitors that lost most of their envelope, either due to high initial mass or very fast rotation.
\label{fig:ET_REMej}}
\end{figure}

We first measure the coefficient of the scaling given in equation \ref{eq:ETscaling}, in our progenitor sample.
As can be seen in figure \ref{fig:ET_REMej} for progenitors that retain most of their envelope, $\Menv/\Mej>0.6$, the relation   
\begin{multline}\label{eq:ET_ERMej}
\Et \approx 0.15 \Eexp^{1/2}R_*\Mej^{1/2} \\ = 0.2 \frac{\Eexp R_*}{\vej} =  0.1 \vej R_* \Mej
\end{multline}
is accurate to within about 30\%. Such progenitors include almost all the models with initial mass $\Mi \leq 20~M_\odot$ that have a massive envelope  ($\Menv>3~M_\odot$). Pre-explosion images of progenitors and light curve modeling of type II-P SNe \citep[e.g.,][ and references therein]{Smartt15} suggest that at least progenitors of this SN type fall into this category. Therefore equation \ref{eq:ET_ERMej} can be applied within fair accuracy to regular type II-P SNe. 

\begin{figure}[t!]
\includegraphics[width=\linewidth]{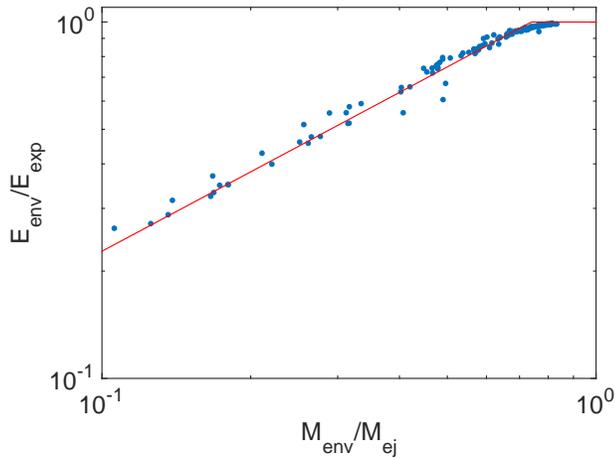}
\caption{The fraction of the energy carried by the envelope out of the total explosion energy, $\Eenv/\Eexp$, as a function of  $\Menv/\Mej$. The solid line is $min\{1.25(\Menv/\Mej)^{0.74},1\}$.
\label{fig:EenvEexp}}
\end{figure}

Figure \ref{fig:ET_REMej} also shows that although equation \ref{eq:ET_ERMej} is rather accurate for progenitors with $\Menv/\Mej>0.6$ it becomes less accurate as the envelope mass fraction drops, becoming inaccurate by more than an order of magnitude for $\Menv/\Mej<0.2$. To understand why low values of $\Menv/\Mej$ affect the scaling of equation \ref{eq:ET_ERMej} we recall that $\Et$ is proportional to the {\it radiation} energy at the beginning of the homologous phase and not directly to the total explosion energy. In progenitors with low $\Menv/\Mej$ most of the explosion energy is deposited in the core, but all the radiation energy deposited in the core is lost via adiabatic expansion well before the homologous phase. Only radiation energy deposited in the envelope remains by the beginning of the homologous phase. The total energy deposited in the envelope is roughly proportional to $(\Menv/\Mej)^{-0.74}$, as can be seen in figure \ref{fig:EenvEexp}. Thus, a low value of $\Menv/\Mej$ implies that a smaller fraction of the explosion energy contributes to $\Et$.

\begin{figure}[t!]
\includegraphics[width=\linewidth]{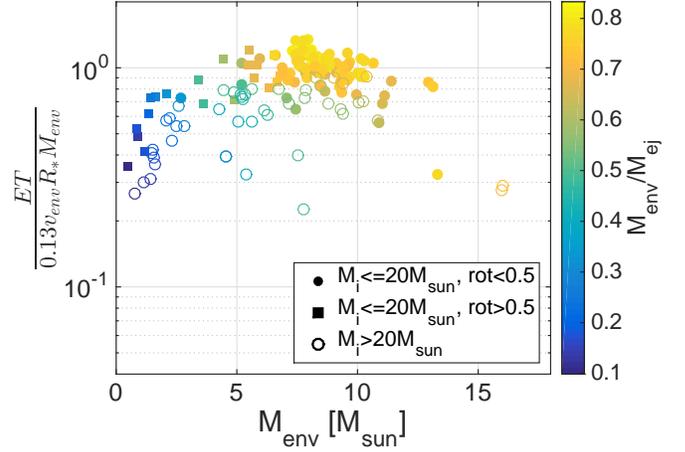}
\caption{Relation between $\Et$ and the estimator $\venv R_* \Menv$. Symbols are the same as in figure \ref{fig:ET_REMej}. This estimator, which depends on envelope properties only, shows a significant improvement over equation \ref{eq:ET_ERMej} for progenitors with low envelope mass, and moderate improvement for progenitors with massive envelopes. Still, for low mass envelope progenitors it is accurate only to within a factor of about 3. 
\label{fig:ET_REMenv}}
\end{figure} 

In fact, the emission during the photospheric phase, after removing \Ni\ contribution, is completely dominated by the radiation energy deposited in the envelope during the SN explosion (hence the term `cooling envelope emission'). Thus, 
$\Et$, is actually a direct probe of the envelope properties and is not directly sensitive to properties of the core such as its mass or velocity. The same is true for any other probe that depends mostly on the cooling envelope emission (such as the photospheric velocity during the plateau, the plateau luminosity and duration, etc.). Therefore it is most useful to define scaling of $\Et$ that depends only on envelope properties:   
\begin{multline}\label{eq:ET_vRMenv}
\Et \approx 0.18 \Eenv^{1/2}R_*\Menv^{1/2} \\ = 0.26 \frac{\Eenv R_*}{\venv}=0.13 \venv R_* \Menv.
\end{multline}
Here, $\Eenv$ is the energy carried by the envelope to infinity and $\venv=\sqrt{2\Eenv/\Menv}$ is the envelope (mass weighted) RMS velocity. Figure \ref{fig:ET_REMenv} shows that indeed $\Et$ provides better estimates for global envelope properties than to those of the entire ejecta. Equation \ref{eq:ET_vRMenv} provides a moderate improvement to progenitors that retain most of their envelope ($\Menv/\Mej>0.6$), where it is accurate to within about 20\% (the normalization of equation \ref{eq:ET_vRMenv} was chosen to better match such progenitors). More importantly, equation \ref{eq:ET_vRMenv} is applicable also to progenitors that lost most of their envelope, where it is accurate to within a factor of about 3.

\begin{figure}[t!]
\includegraphics[width=\linewidth]{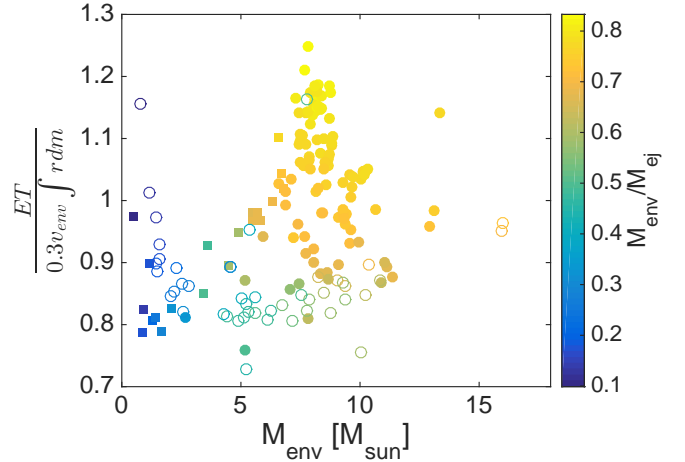}
\caption{Relation between $\Et$ and the estimator $\venv\int r dm$. Symbols are the same as in figure \ref{fig:ET_REMej}. This estimator is accurate to within 20\% for all the progenitors in our sample. 
\label{fig:venv_intrdm}}
\end{figure}

So far we have ignored the density structure of the envelope, which of course also affects the value of $\Et$. When crossing a mass element, the shock deposits half of its energy as kinetic and half as internal (i.e., radiation). The expansion (and adiabatic cooling) of the element begins right after the shock crossing. Thus, the fraction of the deposited energy that remains once the homologous phase starts depends on the initial radius of each mass element and not directly on $R_*$. In both equations \ref{eq:ET_ERMej} and \ref{eq:ET_vRMenv}, $R_*$ is used as the envelope radius. However, explosions of two progenitors with similar $\venv$, $\Menv$, $R_*$ and different density profiles will result in different values of $\Et$. Specifically, $ET$ will be lower in the progenitor that is more concentrated (i.e., most of its envelope mass is at a smaller radius). To account for that, one can attribute the initial radius to each mass element by replacing $\Menv R_*$ with the integral $\int r dm$. Note that the value of this integral is insensitive to whether it is taken over the envelope mass alone or over the entire ejecta since it is completely dominated by mass elements at large radii. Using this integral we find
\begin{multline}\label{eq:ET_vintrdm}
\Et \approx 0.3 ~\venv \int r dm = 0.3~ \venv \Renv \Menv= \\
=0.42~\Eenv^{1/2}\Renv \Menv^{1/2}=0.6~ \frac{\Eenv \Renv}{\venv},
\end{multline}
where  
\begin{equation}
	\Renv=\frac{\int r dm}{\Menv}
\end{equation}
is the mass weighted average radius of the envelope. Equation \ref{eq:ET_vintrdm} is accurate to within about 20\% to all the progenitors in our sample, as can be seen in Figure \ref{fig:venv_intrdm}. Note that while $\int r dm$ hardly changes if the integral is performed over the entire progenitor including the core, $\bar R_{\rm env}$ does change due to the different mass in the denominator of equation 13.

\section{supernovae with spectral measurement of the photospheric velocity}
\label{sec:spectral}
In many SNe spectral measurements provide information about the ejecta velocity. In particular for type II SNe, lines of Fe II and Sc II are considered to be good indicators of the photospheric velocity. In these SNe the photosphere crosses the H envelope from the outside in during the photospheric phase, providing a `scan' of the envelope velocity range. The propagation of the photosphere at early time, before recombination becomes significant, depends only on the density and velocity profile of the ejecta and is relatively simple to model. Recombination becomes significant typically around day 20, when the photosphere has crossed only a very small fraction of the envelope ($<0.1~M_\odot$; see \citealt{Shussman16}). The time at which the photosphere ends crossing the envelope is marked in type II SNe by a sharp drop in the luminosity and is observed typically around day 100. Thus, the photospheric velocity as measured around day 50, typically denoted $v_{50}$, provides a reasonable estimate of $\venv$. Moreover, since the photospheric velocity evolves roughly as $t^{-0.5}$ \citep{Nugent06,Faran14a} it varies between day 20 and 100 by a factor of $\approx 2.2$, implying that $v_{50}$ estimates $\venv$ to an accuracy of 50\% at worst, and most likely much better.

\begin{figure}[t!]
\includegraphics[width=\linewidth]{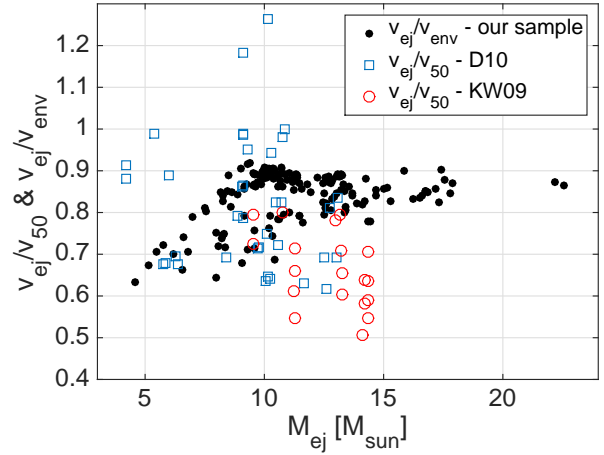}
\caption{The ratio $\vej/v_{50}$ as found in the radiative-hydro simulations presented in \cite{Dessart10} ({\it empty squares}) and \cite{Kasen09} ({\it empty circles}). These are compared to the ratio $\vej/\venv$ found in our numerical progenitors sample ({\it filled circles}).
\label{fig:v_v50}}
\end{figure}

The quality of $v_{50}$ as an estimator of $\venv$ can also be estimated from numerical modeling of type II SN light curves. For that we use the results of \cite{Dessart10} (table 2 therein) and \cite{Kasen09} (table 2 therein). Both studies present a set of hydrodynamic numerical SNe simulations that include detailed radiative transfer to explore properties of various observables including $v_{50}$. Unfortunately, these publications do not include the values of $\venv$, but they do provide the values of $\vej$. Figure \ref{fig:v_v50} depicts $\vej/v_{50}$ in \cite{Dessart10} and \cite{Kasen09}, showing that they find in general $\vej \approx (0.8 \pm 0.2) v_{50}$. It also shows the values of $\vej/\venv$ in our progenitor sample showing $\vej \approx (0.8 \pm 0.1) \venv$. Thus, the reasonable assumption that $\vej/\venv$ in our sample are similar to those of \cite{Dessart10} and \cite{Kasen09}, implies that $v_{50}$ is a good estimator of $\venv$ and that at least in these simulations it is accurate to within about 25\%.

\begin{figure}[t!]
\includegraphics[width=\linewidth]{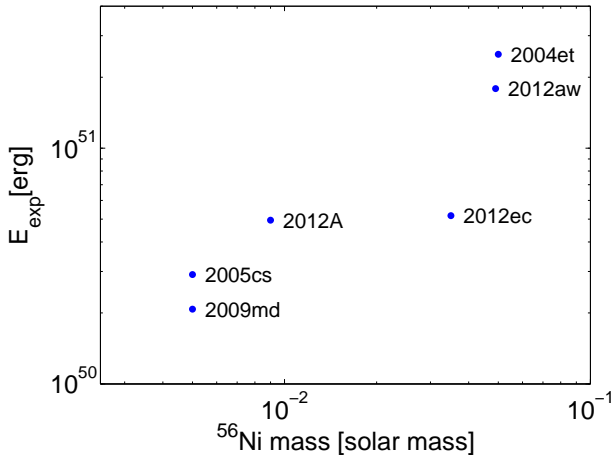}
\caption{The explosion energy of six SNe for which $\Et$, $v_{50}$ and $R_*$ are all measured independently, calculated using equation \ref{eq:ET_ERMej}, plotted as a function of the $^{56}$Ni mass in the explosion.
\label{fig:Eexp_Ni}}
\end{figure}

 \begin{figure}[t!]
\includegraphics[width=\linewidth]{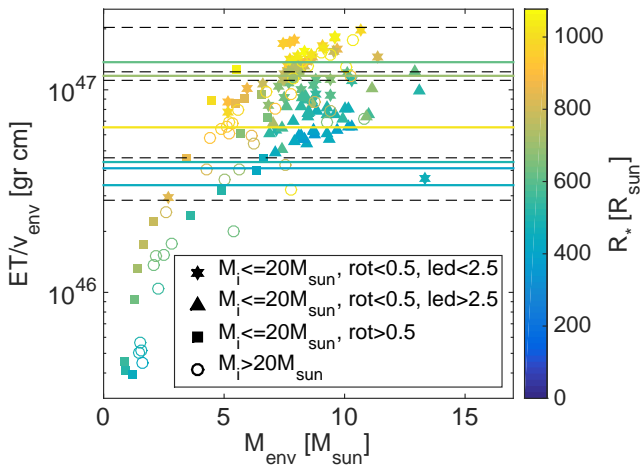}
\caption{$\Et/\venv$ of all the numerically calculated progenitors in our sample as a function of $\Menv$ and color coded according to $R_*$. Progenitors are divided to four groups, based on their initial conditions. {\it Open circles}: massive ($\Mi>20~M_\odot$); {\it Filled rectangles}: fast rotating ($rot>0.5$ of the breakout rotation rate and $\Mi<=20~M_\odot$); {\it filled triangles}: large mixing length parameter ($led>2.5$, rotation lower than 0.5 breakout and $\Mi<=20~M_\odot$); {\it filled hexagons}: small mixing length parameter ($led<2.5$, rotation lower than 0.5 breakout and $\Mi<=20~M_\odot$). Vertical lines are the $\Et/v_{50}$ values of the 11 SNe listed in table \ref{table:ETvR}. Dashed lines for SNe without pre-explosion progenitor images and filled lines for SNe with pre-explosion images. For these the color line is color coded based on the progenitor radius as inferred from its luminosity and temperature in the pre-explosion image.       
\label{fig:ETv}}
\end{figure}

Therefore, in SNe where both $\Et$ and $v_{50}$ are measured we can estimate two interesting quantities based on equations \ref{eq:ET_ERMej} - \ref{eq:ET_vintrdm}. The first is $\Et \cdot v_{50}$ which is an estimator of $\Eexp R_*$ (or more accurately of $\Eenv R_*$). This can be used, for example, to estimate the explosion energy in cases where there are pre-explosion images of the progenitor, which constrain $R_*$. We have done that for six SNe for which $\Et$, $v_{50}$ and $R_*$ are all measured independently (see table \ref{table:ETvR}). Given that in all six the progenitor seems like normal red supergiants with initial masses $\lesssim 15~M_\odot$ \citep{Smartt15}, equation  \ref{eq:ET_ERMej} can probably be used to estimate $\Eexp$ within fair accuracy (within a factor of 2 given the large uncertainty in $R_*$). In figure \ref{fig:Eexp_Ni} we depict $\Eexp$, as estimated based on equation \ref{eq:ET_ERMej} for these six SNe, as a function of the $^{56}$Ni mass in the explosion. Even with that small number of SNe it is clear that $\Eexp$ varies by at least an order of magnitude between various SNe, and that SNe with $\Eexp \sim 2 \times 10^{50}$ erg are common (these SNe are also fainter and thus harder to detect, so their fraction of the total SNe volumetric rate is probably higher than their representation in the observed sample). In addition, a strong, roughly linear, correlation exists between $\Eexp$ and the $^{56}$Ni mass \citep{Kushnir15}. This correlation is most likely the main source of the well known correlation between the plateau luminosity, $v_{50}$ and the $^{56}$Ni mass \citep{Hamuy03}.

The second interesting quantity is $\Et/ v_{50}$ which is an estimator of $\Et/\venv \propto \Menv R_*$. The main 
advantage of $\Et/\venv$ is that it is a property of the progenitor, namely it depends only on the progenitor 
structure, with no dependence on the explosion energy. Thus, we can compare the observed values of $\Et/ 
v_{50}$ to the values of $\Et/\venv$ in our sample of numerical progenitors. This comparison is shown in 
figure \ref{fig:ETv}, which depicts the observed values of the 11 SNe listed in table \ref{table:ETvR} as horizontal
lines, solid for SNe with pre-explosion progenitor images and dashed for SNe without. It also shows all the values for
our numerical progenitors, which are divided to four groups, each marked differently, according to the initial conditions. It can be seen that numerical progenitors that lost most of their envelope ($\Menv<4~M_\odot$) have $\Et/\venv$ that is lower than the observed ones. Almost all other numerical progenitors have $\Et/\venv$ within the observed range, but many are in the high end 
of that range ($1.5-2 \cdot 10^{47} $ gr cm) and few are in the low end ($0.25-0.5 \cdot 10^{47} $ gr cm). This is significant since half of the observed SNe in the sample have low values of $\Et/\venv$ and these are also the fainter ones, implying that they are most likely more abundant than their relative fraction in the sample. 

A lower value of $\Et/\venv$ implies a lower value of $\Menv R_*$. In our numerical sample there are two groups
for which $\Menv R_*$ is relatively small. One is the group of stars with small $\Menv$, which had lost a significant
fraction of their envelope, either via fast rotation ($> 50\%$ of breakout velocity at birth) or due to a high
initial mass ($>20 M_\odot$). However, this group does not seem to be the main origin of the observed low $\Et/\venv$ SNe. The reason for that is that high initial mass progenitors are not common (in all the cases where a progenitor was observed directly its mass was $<20 M_\odot$; \citealt{Smartt15}). Rotation, on the other hand, seems to need a fine tuning in order to lose just the correct amount of envelope mass (most of the fast rotating models lost too much of their mass while others did not lose enough). The second group of low $\Et/\venv$ progenitors are stars with smaller $R_*$ ($\sim 400~R_\odot$) due to a larger value of the mixing length parameter ($3$ or $5$). Given the uncertainty in the value of this coefficient we consider this option as the more likely explanation for the large number of low $\Et/\venv$ SNe. This option is also supported by the measured radii of progenitors observed in pre-explosion images, where the three SNe with $\Et/\venv <0.5$ in our sample with pre-explosion images have $R_* \approx 500~R_\odot$. The result that many SNe progenitors have radii that are $\lesssim 500 ~R_\odot$ is also supported by other, independent lines of evidence \citep{Dessart13,Davies2013,Gall15,Gonzalez15}

\begin{deluxetable*}{p{1cm}cccccc}
\tabletypesize{\scriptsize}
\tablecolumns{6} 
\tablewidth{0pt}
\setlength{\tabcolsep}{0.0in}
\tablecaption{SNe with measured $\Et$ and $v_{50}$}
\tablehead{
SN &$\Et$ & $v_{50}$ & $L_{25}$ & $R_*^\dagger$ &Ref.\\
    &[$10^{55}{\rm erg\,s}$] & [km/s] &[$10^{41}{\rm erg/s}$] &[$R_\odot$]&}
\startdata  
1999em 		& 4 	& 3280 & 14	  & 	&  1\\
1999gi 		& 4.1 	& 3700 & 16	  & 	& 1,2\\
2004et 		& 5.5 	& 4020 & 23   & 635	& 3\\
2005cs 		& 0.86	& 1980 & 1.9  & 420	&   4\\
2007od 		& 6.4 	& 3130 & 32   & 	& 5\\
2009N 		& 1.2 	& 2580 & 4.6  &		&  6\\
2009ib      & 0.88 	& 3090 & 5.2  & 	&   7\\
2009md 		& 0.68 	& 2000 & 2.4  & 471	& 8\\
2012A 		& 1.2 	& 2840 & 6.1  & 516	& 9\\
2012aw 		& 4.6 	& 3890 & 16   & 713	& 10\\
2012ec 	    & 2.2  	& 3370 & 10   &1031	& 11\\ 
\enddata
 \tablenotetext{}{1. \cite{Bersten09} 2. \cite{Leonard02} 3. \cite{Maguire10} 4. \cite{Pastorello09} 5. \cite{Inserra11} 6. \cite{Takats14} 7. \cite{Takats15}
8. \cite{Fraser11} 9. \cite{Tomasella13} 10. \cite{DallOra14} 11. \cite{Barbarino15} } 
\tablenotetext{$^\dagger$}{Progenitor radius based on luminosity and effective temperature measured in pre-explosion images and listed in \cite{Smartt15}. The reported errors in the measure luminosities and effective temperatures translate to errors of about 25\%-50\% in the radii.}
\tablenotetext{}{A list of SNe with measured $\Et$ and $v_{50}$ taken from \cite{Nakar15}. The light curves used to measure $\Et$ are reported in the listed references.}
\label{table:ETvR}
\end{deluxetable*}

\section{Conclusions}
\label{sec:Conclusions}
We have used a large set of 145 numerical stars, calculated to the point of core collapse using the stellar evolution code MESA, in order to study the relation between the observable $\Et$ and SN progenitors. $\Et$ can be measured for any SN with observed bolometric light curve. It is a time weighted integral of the cooling envelope emission, with radioactive contribution removed, and it measures the amount of radiation energy that is deposited in the envelope at the beginning of the homologous phase. As such it constrains directly the structure of the progenitor and the explosion energy, without any dependence on radiation transfer physics or on the contribution of radioactive decay to the observed light. 

We have found that although $\Et$ depends on the exact structure of the progenitor, it is an accurate probe of $\Eexp^{1/2}R_*\Mej^{1/2}$ for progenitors that retain most of the hydrogen envelope (equation \ref{eq:ET_ERMej}). Specifically in our progenitor sample it is accurate to within $30\%$ for progenitors with $\Menv/\Mej>0.6$. However, it is not an accurate probe of $\Eexp$ and $\Mej$ for progenitors that lost most of their envelope, since $\Et$ is sensitive only to the fraction of the total energy that is deposited in the envelope, and this fraction is highly dependent on $\Menv/\Mej$. We note that any observable which depends on the photospheric phase of type II SNe (e.g., plateau luminosity and duration), is sensitive mostly to envelope properties and is therefore, similarly to $\Et$, only an approximate probe of the entire ejecta properties.

We have then explored how accurately $\Et$ can be related to envelope properties and found that it is a good probe of $\Eenv^{1/2}R_*\Menv^{1/2} \propto \venv R_* \Menv$ (equation \ref{eq:ET_vRMenv}). In our progenitor sample it is accurate to within $20\%$ for progenitors with $\Menv/\Mej>0.6$ and up to a factor of three for progenitors that lost most of their envelope. The dependence on the envelope mass arises from the fact that progenitors that lost different fractions of their envelope mass have different density structures. We therefore consider a third combination of properties which takes the envelope structure into account, $\venv \Renv \Menv$, where $\Renv$ is the mass weighted average radius of the envelope. In our sample $\Et$ measures $\venv \Renv \Menv$ to within $20\%$ accuracy for all progenitors, regardless of their envelope mass.

When the bolometric light curve is accompanied by spectral measurements of line velocities then both $\Et$ and $\venv$ can be estimated (as we have showed, $v_{50}$ provides a good estimate for $\venv$). 
This is very useful. First, the product $\Et \cdot v_{50}$ is a probe of $\Eexp R_*$. We have used that in stars where $R_*$ was measured from pre-explosion images to estimate $\Eexp$ in a robust way which is insensitive to detailed and uncertain light curve modeling. Second, the ratio $\Et/v_{50}$ depends on the progenitor structure only, and is independent of the explosion energy. We have compared this quantity as measured in 11 SNe to that of our sample of numerical progenitors. This comparison suggests that SNe progenitors often have a radius $R_* \lesssim 500~R_\odot$, which is smaller than the typical RSG radius obtained by stellar evolution models. This result supports previous studies that got to the same conclusion based on completely different arguments \citep{Dessart10,Davies2013,Gonzalez15,Gall15}. Our simulations suggest that the small radii hint to relatively large values (3-5) of the mixing length parameter, as models with these values produce SNe with $\Et/\venv$ that is more similar to the observations.

\acknowledgments We thank Dovi Poznanski for useful comments.  This research was partially supported by the I-CORE Program (1829/12). TS and EN were partially supported by an ERC starting grant (GRB/SN), ISF grant (1277/13) and an ISA grant. BK was partially supported by the Beracha Foundation.

\appendix

The stellar evolution of the progenitor models was followed using the publicly available package MESA version 6596 \citep{2011ApJS..192....3P,2013ApJS..208....4P,2015ApJS..220...15P}. To produce a wide range of progenitors, we varied the zero age main sequence (ZAMS) mass between $[10, 50] M_{\sun}$, the metallicity between $[2 \times 10^{-5}, 2 \times 10^{-2}]$, the mixing length parameter between $[1.5,5]$, and the initial rotation rate between $[0,0.8]$ of the breakup rotation rate. In all models, mass loss was determined according to the "Dutch" recipe in MESA, combining the rates from \cite{2009A&A...497..255G, 1990A&A...231..134N, 2000A&A...360..227N, 2001A&A...369..574V}, with a coefficient $\eta=1$, the convection was according to the Ledoux criterion, with a semi-convection efficiency parameter $\alpha_{sc}=0.1$ \citep[eq. 12]{2013ApJS..208....4P}, and exponential overshoot with parameter $f=0.008$ \citep[eq. 2]{2011ApJS..192....3P}.

The properties of the numerical progenitors are summarised in table \ref{table:resultsb}. The initial (ZAMS) parameters are shown in rows 1-4, while the properties just before to the core collapse appear in rows 5-9. The key features of the explosion are given in rows 10-11, and are normalized to be independent of the explosion energy. We denote $E_{\rm{exp},51}=E_{\rm{exp}}/10^{51}$erg.

\begin{longtable*}{ccccccccccc}
\tablecaption{Initial parameters and properties of the numerical progenitors}
\tablehead{
$M_{ZAMS}$ & $Z$ & mixing length & rotation & $M_{\rm{final}}$ & $\Mej$ & $\Menv$ & $R_*$ & $\Renv$ & $\Eenv/\Eexp$ & $\Et / E_{\rm{exp},51}^{1/2}$ \\
$[M_{\odot}]$ & & parameter & [breakup] & $[M_{\odot}]$ & $[M_{\odot}]$ & $[M_{\odot}]$ & $[R_{\odot}]$ & $[R_{\odot}]$ & & $[10^{55}\rm erg~s]$
}
11 & 0.02 & 2 & 0.4 & 10.07 & 9.72 & 7.19 & 518 & 225 & 0.97 & 2.34 \\ 
12 & 0.0002 & 1.5 & 0 & 12.94 & 11.39 & 8.82 & 608 & 280 & 0.98 & 3.76 \\ 
12 & 0.0002 & 1.5 & 0.2 & 12.89 & 11.27 & 8.72 & 603 & 271 & 0.98 & 3.56 \\ 
12 & 0.0002 & 2 & 0 & 12.93 & 11.31 & 8.80 & 553 & 237 & 0.98 & 3.19 \\ 
12 & 0.0002 & 2 & 0.2 & 12.88 & 11.28 & 8.68 & 552 & 230 & 0.98 & 2.99 \\ 
12 & 0.0002 & 2 & 0.4 & 12.89 & 11.19 & 8.25 & 609 & 262 & 0.97 & 3.16 \\ 
12 & 0.0002 & 3 & 0 & 12.94 & 11.31 & 8.85 & 480 & 193 & 0.98 & 2.63 \\ 
12 & 0.0002 & 3 & 0.2 & 12.93 & 11.25 & 8.54 & 512 & 205 & 0.97 & 2.63 \\ 
12 & 0.0002 & 3 & 0.4 & 12.89 & 11.25 & 8.46 & 511 & 200 & 0.97 & 2.51 \\ 
12 & 0.002 & 1.5 & 0 & 11.59 & 10.02 & 7.70 & 736 & 353 & 0.98 & 4.43 \\ 
12 & 0.002 & 1.5 & 0.2 & 10.49 & 8.86 & 5.94 & 867 & 438 & 0.95 & 4.09 \\ 
12 & 0.002 & 1.5 & 0.4 & 10.80 & 9.17 & 6.60 & 783 & 379 & 0.97 & 4.10 \\ 
12 & 0.002 & 2 & 0 & 12.23 & 10.59 & 8.20 & 614 & 270 & 0.98 & 3.51 \\ 
12 & 0.002 & 2 & 0.2 & 11.60 & 9.94 & 7.49 & 633 & 279 & 0.97 & 3.35 \\ 
12 & 0.002 & 2 & 0.4 & 11.32 & 9.64 & 6.86 & 681 & 298 & 0.96 & 3.16 \\ 
12 & 0.002 & 3 & 0 & 12.48 & 10.85 & 8.47 & 489 & 198 & 0.98 & 2.63 \\ 
12 & 0.002 & 3 & 0.2 & 11.98 & 10.41 & 7.85 & 502 & 201 & 0.97 & 2.48 \\ 
12 & 0.002 & 3 & 0.4 & 11.25 & 9.60 & 6.89 & 532 & 210 & 0.96 & 2.29 \\ 
12 & 0.002 & 5 & 0 & 12.66 & 11.02 & 8.72 & 396 & 153 & 0.99 & 2.12 \\ 
12 & 0.002 & 5 & 0.2 & 11.48 & 9.86 & 7.12 & 438 & 161 & 0.96 & 1.82 \\ 
12 & 0.002 & 5 & 0.4 & 12.23 & 10.65 & 7.90 & 432 & 161 & 0.97 & 1.96 \\ 
12 & 0.02 & 1.5 & 0 & 11.51 & 9.97 & 7.93 & 910 & 454 & 0.98 & 6.22 \\ 
12 & 0.02 & 1.5 & 0.2 & 11.36 & 9.93 & 7.74 & 926 & 466 & 0.98 & 6.13 \\ 
12 & 0.02 & 1.5 & 0.4 & 11.16 & 9.59 & 7.45 & 948 & 481 & 0.98 & 6.20 \\ 
12 & 0.02 & 1.5 & 0.6 & 9.93 & 8.26 & 5.52 & 1052 & 563 & 0.95 & 5.27 \\ 
12 & 0.02 & 2 & 0 & 11.70 & 10.07 & 8.08 & 709 & 319 & 0.98 & 4.30 \\ 
12 & 0.02 & 2 & 0.2 & 11.49 & 9.92 & 7.84 & 752 & 337 & 0.98 & 4.47 \\ 
12 & 0.02 & 2 & 0.4 & 11.23 & 9.65 & 7.48 & 778 & 350 & 0.98 & 4.38 \\ 
12 & 0.02 & 2 & 0.6 & 10.29 & 8.65 & 5.84 & 841 & 387 & 0.94 & 3.66 \\ 
12 & 0.02 & 2 & 0.8 & 7.41 & 5.49 & 1.41 & 714 & 278 & 0.52 & 0.80 \\ 
12 & 0.02 & 3 & 0 & 11.92 & 10.34 & 8.36 & 559 & 229 & 0.99 & 3.22 \\ 
12 & 0.02 & 3 & 0.2 & 11.86 & 10.27 & 8.23 & 568 & 232 & 0.99 & 3.19 \\ 
12 & 0.02 & 3 & 0.4 & 11.41 & 9.81 & 7.60 & 592 & 238 & 0.98 & 3.00 \\ 
12 & 0.02 & 3 & 0.6 & 11.19 & 9.53 & 6.78 & 650 & 253 & 0.95 & 2.74 \\ 
12 & 0.02 & 3 & 0.8 & 7.16 & 5.17 & 0.87 & 553 & 160 & 0.37 & 0.30 \\ 
12 & 0.02 & 5 & 0 & 12.27 & 10.70 & 8.70 & 383 & 149 & 0.99 & 2.16 \\ 
12 & 0.02 & 5 & 0.2 & 11.87 & 10.31 & 8.19 & 388 & 148 & 0.99 & 2.07 \\ 
12 & 0.02 & 5 & 0.4 & 11.78 & 10.16 & 8.11 & 390 & 149 & 0.99 & 2.08 \\ 
12 & 0.02 & 5 & 0.6 & 11.00 & 9.27 & 6.66 & 445 & 159 & 0.96 & 1.75 \\ 
12 & 0.02 & 5 & 0.8 & 10.08 & 8.27 & 4.87 & 510 & 166 & 0.86 & 1.35 \\ 
13 & 0.02 & 2 & 0 & 11.71 & 10.00 & 8.11 & 708 & 318 & 0.98 & 4.37 \\ 
13 & 0.02 & 2 & 0.2 & 11.56 & 9.95 & 7.91 & 711 & 318 & 0.98 & 4.17 \\ 
13 & 0.02 & 2 & 0.4 & 11.26 & 9.53 & 7.48 & 739 & 333 & 0.98 & 4.13 \\ 
13 & 0.02 & 2 & 0.6 & 10.01 & 8.18 & 5.50 & 847 & 389 & 0.93 & 3.56 \\ 
14 & 0.02 & 2 & 0 & 12.34 & 10.68 & 8.33 & 780 & 354 & 0.98 & 4.55 \\ 
14 & 0.02 & 2 & 0.2 & 12.04 & 10.28 & 7.89 & 817 & 375 & 0.97 & 4.66 \\ 
14 & 0.02 & 2 & 0.6 & 8.94 & 6.98 & 3.41 & 851 & 380 & 0.79 & 2.21 \\ 
15 & 2e-05 & 1.5 & 0 & 14.98 & 13.27 & 10.29 & 555 & 247 & 0.98 & 3.44 \\ 
15 & 2e-05 & 1.5 & 0.4 & 14.76 & 12.74 & 9.40 & 601 & 250 & 0.96 & 3.03 \\ 
15 & 2e-05 & 3 & 0 & 14.97 & 13.07 & 10.18 & 465 & 182 & 0.98 & 2.51 \\ 
15 & 2e-05 & 3 & 0.4 & 14.91 & 12.77 & 9.59 & 496 & 188 & 0.97 & 2.41 \\ 
15 & 2e-05 & 5 & 0 & 14.98 & 13.22 & 10.30 & 390 & 146 & 0.98 & 2.05 \\ 
15 & 2e-05 & 5 & 0.4 & 14.90 & 12.85 & 9.35 & 442 & 152 & 0.95 & 1.84 \\ 
15 & 0.0002 & 1.5 & 0 & 14.95 & 13.10 & 10.13 & 608 & 277 & 0.98 & 3.82 \\ 
15 & 0.0002 & 1.5 & 0.4 & 14.77 & 12.70 & 9.40 & 658 & 295 & 0.97 & 3.65 \\ 
15 & 0.0002 & 3 & 0 & 14.92 & 13.02 & 10.02 & 476 & 188 & 0.98 & 2.55 \\ 
15 & 0.0002 & 3 & 0.4 & 14.83 & 12.79 & 9.29 & 524 & 201 & 0.96 & 2.46 \\ 
15 & 0.0002 & 5 & 0 & 14.93 & 13.16 & 10.14 & 394 & 149 & 0.98 & 2.06 \\ 
15 & 0.0002 & 5 & 0.4 & 14.79 & 12.60 & 9.27 & 448 & 161 & 0.96 & 1.97 \\ 
15 & 0.002 & 1.5 & 0 & 14.27 & 12.53 & 9.56 & 778 & 376 & 0.98 & 5.02 \\ 
15 & 0.002 & 1.5 & 0.4 & 10.37 & 8.62 & 5.18 & 894 & 466 & 0.91 & 3.69 \\ 
15 & 0.002 & 3 & 0 & 14.10 & 12.21 & 9.28 & 518 & 207 & 0.97 & 2.68 \\ 
15 & 0.002 & 3 & 0.4 & 12.62 & 10.58 & 7.41 & 562 & 219 & 0.94 & 2.33 \\ 
15 & 0.002 & 5 & 0 & 14.44 & 12.68 & 9.69 & 401 & 152 & 0.98 & 2.03 \\ 
15 & 0.002 & 5 & 0.4 & 13.63 & 11.79 & 8.45 & 432 & 158 & 0.95 & 1.84 \\ 
15 & 0.02 & 2 & 0 & 13.05 & 11.27 & 8.68 & 835 & 382 & 0.97 & 4.91 \\ 
15 & 0.02 & 2 & 0.2 & 12.66 & 10.94 & 8.17 & 841 & 383 & 0.97 & 4.57 \\ 
15 & 0.02 & 2 & 0.4 & 13.15 & 11.37 & 8.60 & 845 & 385 & 0.97 & 4.77 \\ 
15 & 0.02 & 2 & 0.6 & 7.84 & 5.77 & 1.67 & 773 & 314 & 0.56 & 1.00 \\ 
16 & 0.02 & 2 & 0 & 14.47 & 12.68 & 9.67 & 851 & 385 & 0.97 & 5.05 \\ 
16 & 0.02 & 2 & 0.2 & 13.25 & 11.47 & 8.38 & 883 & 404 & 0.96 & 4.70 \\ 
16 & 0.02 & 2 & 0.4 & 12.71 & 10.78 & 7.66 & 943 & 440 & 0.96 & 4.69 \\ 
16 & 0.02 & 2 & 0.6 & 8.62 & 6.62 & 2.10 & 777 & 313 & 0.58 & 1.19 \\ 
17 & 0.02 & 2 & 0 & 14.25 & 12.47 & 9.10 & 963 & 451 & 0.96 & 5.43 \\ 
17 & 0.02 & 2 & 0.2 & 13.44 & 11.59 & 8.07 & 977 & 456 & 0.95 & 4.87 \\ 
17 & 0.02 & 2 & 0.4 & 13.19 & 11.25 & 7.73 & 1009 & 480 & 0.94 & 4.96 \\ 
17 & 0.02 & 2 & 0.6 & 8.50 & 6.20 & 1.31 & 625 & 208 & 0.43 & 0.53 \\ 
18 & 0.02 & 2 & 0 & 16.25 & 14.23 & 10.67 & 978 & 451 & 0.97 & 5.98 \\ 
18 & 0.02 & 2 & 0.2 & 15.22 & 13.32 & 9.61 & 1015 & 478 & 0.96 & 5.78 \\ 
18 & 0.02 & 2 & 0.4 & 13.89 & 11.93 & 8.06 & 1031 & 489 & 0.94 & 5.06 \\ 
18 & 0.02 & 2 & 0.6 & 8.58 & 6.57 & 0.91 & 512 & 131 & 0.32 & 0.24 \\ 
19 & 0.02 & 2 & 0 & 15.46 & 13.53 & 9.58 & 1046 & 491 & 0.95 & 5.78 \\ 
19 & 0.02 & 2 & 0.2 & 14.47 & 12.50 & 8.34 & 1077 & 509 & 0.93 & 5.23 \\ 
19 & 0.02 & 2 & 0.6 & 9.42 & 6.78 & 1.17 & 452 & 90 & 0.35 & 0.22 \\ 
20 & 2e-05 & 3 & 0.4 & 17.88 & 17.17 & 10.91 & 666 & 187 & 0.87 & 2.08 \\ 
20 & 0.0002 & 1.5 & 0 & 19.94 & 17.41 & 13.33 & 471 & 58 & 0.94 & 0.98 \\ 
20 & 0.0002 & 1.5 & 0.4 & 19.48 & 16.86 & 11.38 & 812 & 349 & 0.94 & 4.18 \\ 
20 & 0.0002 & 3 & 0 & 19.91 & 17.90 & 12.93 & 616 & 238 & 0.96 & 3.35 \\ 
20 & 0.0002 & 3 & 0.4 & 19.45 & 16.76 & 11.12 & 680 & 245 & 0.92 & 2.92 \\ 
20 & 0.0002 & 5 & 0 & 19.92 & 17.57 & 13.11 & 511 & 185 & 0.97 & 2.71 \\ 
20 & 0.0002 & 5 & 0.4 & 19.38 & 16.76 & 11.02 & 561 & 184 & 0.91 & 2.19 \\ 
20 & 0.002 & 1.5 & 0 & 14.69 & 12.62 & 7.83 & 1027 & 522 & 0.92 & 4.74 \\ 
20 & 0.002 & 1.5 & 0.4 & 12.73 & 10.59 & 5.18 & 991 & 472 & 0.80 & 3.05 \\ 
20 & 0.002 & 3 & 0 & 15.70 & 13.67 & 8.65 & 703 & 271 & 0.90 & 2.76 \\ 
20 & 0.002 & 3 & 0.4 & 14.72 & 12.44 & 7.04 & 735 & 271 & 0.83 & 2.36 \\ 
20 & 0.002 & 5 & 0 & 16.70 & 14.69 & 9.94 & 529 & 186 & 0.92 & 2.20 \\ 
20 & 0.02 & 2 & 0 & 15.41 & 13.38 & 9.11 & 1025 & 472 & 0.94 & 5.18 \\ 
20 & 0.02 & 2 & 0.2 & 15.00 & 13.02 & 8.63 & 1019 & 469 & 0.93 & 4.91 \\ 
20 & 0.02 & 2 & 0.4 & 10.10 & 8.00 & 2.68 & 830 & 325 & 0.59 & 1.39 \\ 
21 & 0.02 & 2 & 0 & 15.74 & 13.70 & 9.04 & 1037 & 465 & 0.92 & 4.90 \\ 
21 & 0.02 & 2 & 0.2 & 14.98 & 12.95 & 8.26 & 1030 & 461 & 0.91 & 4.63 \\ 
21 & 0.02 & 2 & 0.4 & 11.47 & 9.18 & 4.43 & 925 & 389 & 0.77 & 2.44 \\ 
22 & 0.02 & 2 & 0 & 17.37 & 15.01 & 10.36 & 1039 & 456 & 0.94 & 5.33 \\ 
22 & 0.02 & 2 & 0.2 & 10.56 & 8.29 & 2.59 & 799 & 283 & 0.56 & 1.17 \\ 
22 & 0.02 & 2 & 0.4 & 12.31 & 10.23 & 4.88 & 945 & 392 & 0.76 & 2.55 \\ 
22 & 0.02 & 2 & 0.6 & 10.73 & 8.07 & 1.45 & 473 & 92 & 0.35 & 0.25 \\ 
23 & 0.02 & 2 & 0 & 16.80 & 14.70 & 9.40 & 1032 & 442 & 0.90 & 4.65 \\ 
23 & 0.02 & 2 & 0.2 & 13.09 & 11.04 & 5.60 & 986 & 417 & 0.79 & 3.01 \\ 
23 & 0.02 & 2 & 0.4 & 12.94 & 10.80 & 5.13 & 942 & 380 & 0.75 & 2.54 \\ 
23 & 0.02 & 2 & 0.6 & 10.49 & 7.96 & 0.79 & 153 & 15 & 0.24 & 0.03 \\ 
24 & 0.02 & 2 & 0 & 16.32 & 14.22 & 8.47 & 1040 & 440 & 0.87 & 4.26 \\ 
24 & 0.02 & 2 & 0.2 & 14.63 & 12.55 & 6.75 & 1005 & 418 & 0.82 & 3.42 \\ 
24 & 0.02 & 2 & 0.4 & 10.67 & 8.24 & 2.07 & 634 & 187 & 0.46 & 0.65 \\ 
25 & 2e-05 & 1.5 & 0 & 24.97 & 22.17 & 16.01 & 154 & 20 & 0.95 & 0.31 \\ 
25 & 2e-05 & 1.5 & 0.4 & 19.00 & 15.87 & 7.78 & 1018 & 86 & 0.61 & 0.91 \\ 
25 & 0.0002 & 3 & 0.4 & 18.33 & 15.18 & 7.53 & 784 & 160 & 0.67 & 1.28 \\ 
25 & 0.0002 & 5 & 0 & 24.90 & 22.58 & 15.95 & 149 & 19 & 0.94 & 0.29 \\ 
25 & 0.0002 & 5 & 0.4 & 16.20 & 13.22 & 5.38 & 639 & 94 & 0.56 & 0.65 \\ 
25 & 0.002 & 1.5 & 0 & 14.30 & 11.63 & 5.21 & 994 & 436 & 0.74 & 2.61 \\ 
25 & 0.002 & 1.5 & 0.4 & 20.00 & 17.31 & 10.06 & 1030 & 409 & 0.85 & 3.77 \\ 
25 & 0.002 & 3 & 0 & 17.94 & 15.31 & 8.78 & 799 & 292 & 0.84 & 2.71 \\ 
25 & 0.002 & 3 & 0.4 & 19.40 & 16.51 & 9.39 & 756 & 240 & 0.82 & 2.33 \\ 
25 & 0.002 & 5 & 0 & 18.44 & 16.22 & 9.29 & 646 & 208 & 0.83 & 2.09 \\ 
25 & 0.002 & 5 & 0.4 & 20.54 & 17.79 & 10.83 & 640 & 183 & 0.85 & 2.01 \\ 
25 & 0.02 & 2 & 0 & 16.08 & 13.89 & 7.86 & 1016 & 416 & 0.84 & 3.76 \\ 
25 & 0.02 & 2 & 0.2 & 13.62 & 11.46 & 5.34 & 950 & 378 & 0.74 & 2.59 \\ 
25 & 0.02 & 2 & 0.4 & 11.03 & 8.27 & 2.19 & 649 & 194 & 0.48 & 0.71 \\ 
25 & 0.02 & 2 & 0.6 & 11.60 & 9.05 & 1.14 & 260 & 34 & 0.27 & 0.08 \\ 
26 & 0.02 & 2 & 0 & 16.45 & 14.10 & 7.79 & 943 & 360 & 0.82 & 3.14 \\ 
26 & 0.02 & 2 & 0.2 & 13.88 & 11.49 & 5.22 & 900 & 338 & 0.72 & 2.29 \\ 
26 & 0.02 & 2 & 0.4 & 11.10 & 8.38 & 1.51 & 495 & 103 & 0.35 & 0.28 \\ 
27 & 0.02 & 2 & 0 & 16.22 & 13.38 & 7.14 & 917 & 337 & 0.80 & 2.71 \\ 
27 & 0.02 & 2 & 0.2 & 13.37 & 10.61 & 4.28 & 811 & 278 & 0.66 & 1.59 \\ 
27 & 0.02 & 2 & 0.4 & 12.36 & 9.59 & 2.51 & 625 & 169 & 0.46 & 0.65 \\ 
28 & 0.02 & 2 & 0 & 15.66 & 12.81 & 6.12 & 844 & 288 & 0.74 & 2.07 \\ 
28 & 0.02 & 2 & 0.2 & 19.53 & 16.63 & 10.24 & 902 & 325 & 0.87 & 3.44 \\ 
28 & 0.02 & 2 & 0.4 & 11.86 & 9.30 & 1.57 & 467 & 87 & 0.33 & 0.24 \\ 
29 & 0.02 & 2 & 0 & 16.21 & 13.40 & 6.25 & 790 & 254 & 0.72 & 1.85 \\ 
29 & 0.02 & 2 & 0.2 & 12.94 & 10.22 & 2.83 & 638 & 173 & 0.48 & 0.72 \\ 
29 & 0.02 & 2 & 0.4 & 12.94 & 10.36 & 2.29 & 548 & 124 & 0.40 & 0.44 \\ 
30 & 0.02 & 2 & 0 & 16.05 & 13.37 & 5.62 & 703 & 205 & 0.66 & 1.39 \\ 
30 & 0.02 & 2 & 0.2 & 15.34 & 12.52 & 5.04 & 697 & 203 & 0.64 & 1.28 \\ 
30 & 0.02 & 2 & 0.4 & 12.94 & 10.45 & 1.42 & 275 & 38 & 0.29 & 0.10 \\ 
35 & 0.02 & 2 & 0 & 17.11 & 14.47 & 4.55 & 377 & 72 & 0.52 & 0.41 \\ 
35 & 0.02 & 2 & 0.2 & 17.10 & 14.37 & 4.55 & 380 & 73 & 0.52 & 0.42
\label{table:resultsb}
\end{longtable*}

\bibliographystyle{apj}
\bibliography{ms}

\end{document}